# Single- and few-layer graphene growth on stainless steel substrates by direct thermal chemical vapor deposition

Robin John[a§], A Ashokreddy[b§], C Vijayan[a] and T Pradeep[b*]

[a] Department of Physics, Indian Institute of Technology Madras, Chennai – 600 036, India.

[b] DST Unit of Nanoscience (DST UNS), Department of Chemistry, Indian Institute of Technology Madras, Chennai – 600 036, India.

[§] *Equally contributing authors*

## Abstract

Steeping interest on graphene research in basic sciences and applications emphasizes the need for an economical means of synthesizing it. We report a method for the synthesis of graphene on commercially available stainless steel foils using direct thermal chemical vapor deposition. Our method of synthesis and the use of relatively cheap precursors such as ethanol ($CH_3CH_2OH$) as a source of carbon and SS 304 as the substrate, proved to be economically viable. Presence of single- and few-layer graphene was confirmed using confocal Raman microscopy/spectroscopy. X-ray photoelectron spectroscopic measurements were further used to establish the influence of various elemental species present in stainless steel on graphene growth. Role of cooling rate on surface migration of certain chemical species (oxides of Fe, Cr and Mn) that promote or hinder the growth of graphene is probed. Such analysis of the chemical species present on the surface can be promising for graphene based catalytic research.

**Introduction**

One of the most expanding disciplines of contemporary research is the science of graphene. Graphene is a flat single-layer of $sp^2$ bonded carbon atoms that are tightly packed into a two-dimensional (2D) honeycomb lattice, which is the basic building block for graphitic materials of all other dimensionalities, and traditionally used to describe the properties of such structures [1]. Though realized

---

[*] Corresponding author. Fax: + 91-44 2257-0545. E-mail address: pradeep@iitm.ac.in (T. Pradeep)

only in 2004, graphene research took the central stage within a short span of time, owing to its exotic physical properties [2]. Graphene has been proven to be a promising material for various applications. Recently, for example, 100-GHz transistors based on epitaxial graphene [3], hybrid materials with graphene enabled band-gap engineering [4], nanomats for next-generation catalysis and sensing [5], nanomeshs to circumvent the problem of the zero bandgap which hampers its effective usage in electronics [6], heat conductor as well as heat removal quilts in high-power electronics [7], etc. were demonstrated.

Earliest studies used graphene exfoliated mechanically using scotch tape method from highly oriented pyrolitic graphite. Though the yield is very poor for this method, the quality of obtained graphene is high [8]. Influence of graphene, however, is yet to be felt in the commercial market, owing to the lack of mass production methods for the fabrication of high-quality, large-area graphene in usable quantities. Except thermal chemical vapor deposition (CVD), most of the methods developed for the production of graphene are proven to be ill-suited for the commercial-scale production, some being cumbersome and expensive while others resulting in poor or uneven quality of graphene [9]. Various substrates for growing graphene have been used in the recent past and some of these involve the use of an active metallic layer [9, 10]. Most notably, single- and bi-layer graphene coverage of up to 87% area on nickel films [11] and ~95% area single-layer coverage on copper foils [12] were reported. Analogous synthesis methods have been attempted in the case of CNTs and other nano graphitic materials [13-15]. Alcohol based synthesis on catalytically active surfaces is a common synthesis strategy for making organized CNT assemblies [13].

A rapid growth of graphene on simple substrates, such as stainless steel (SS) without elaborate synthetic controls will help in the growth of graphene research. Although graphene synthesis on SS304 substrates has been reported, using microwave plasma CVD and radio frequency plasma enhanced CVD method, the carbon source remains to be $CH_4$ [14-16]. We developed a thermal CVD method for the synthesis of graphene from single- to few-layers, over large surface area using alcohol precursor (ethanol) which in comparison to $CH_4$, economically cheap, easy to be handled and stored. The custom-built horizontal split tube furnace allows us to cool the substrate by opening the furnace. Synthesis was achieved within ~10 minutes after flushing the set up. The grown material was characterized by confocal Raman microscopy/spectroscopy and X-ray photoelectron spectroscopy. Studies reveal that graphene growth is enhanced in specific regions and is retarded in others, depending on the surface chemical composition. This understanding may lead to the development of new catalytic surfaces for uniform graphene growth.

**Experimental**

*Sample preparation*

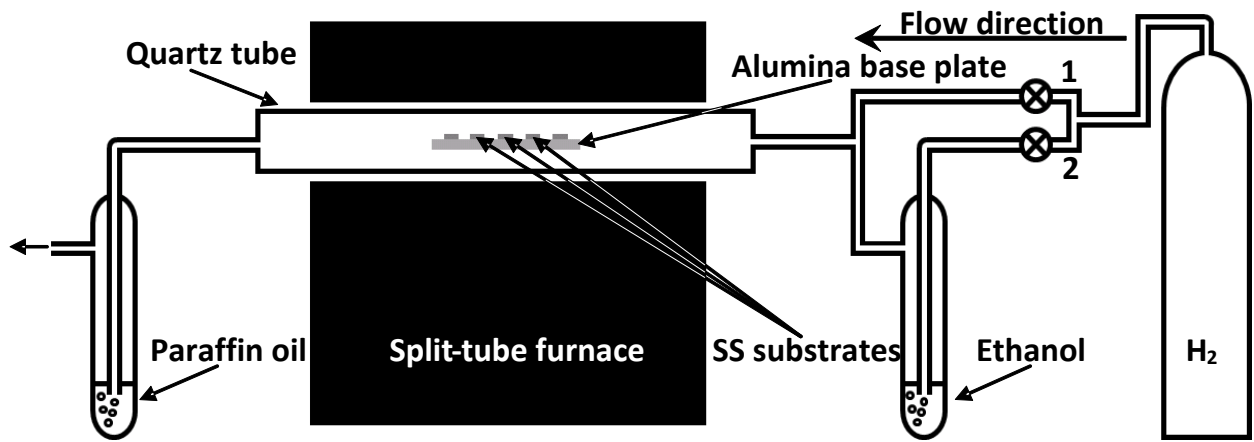

Figure 1. Schematic of the assembled CVD set up, used for the synthesis of graphene. The outlet trap containing paraffin oil acts as a seal to prevent air from entering into the reaction tube.

Schematic of the CVD set-up used for the synthesis of graphene is shown in figure 1. A one-meter long (35-mm ID) quartz tube, horizontally placed inside a hot wall, split tube furnace (designed locally and capable of reaching 1100 ºC), serves as the CVD reactor. The horizontal split tube furnace allows rapid cooling, simply by opening it. The substrate for graphene growth was 0.1 mm thick foil of SS304, cut into a size of ~1 cm × 1 cm, which were supported on an alumina base plate. The reaction chamber was flushed using hydrogen (99.9% purity) at a pressure of little above 1 atm, for 1 h to create a reducing environment. Same flow rate was used throughout the experiment. The SS304 substrates were used as received. Under $H_2$ flow, the temperature of the chamber was increased to the growth temperature, at a heating rate of ~20 ºC/min. At the growth temperature, vapors of spectroscopic purity ethanol (EtOH, Changshu Yungyaun Chemical, (AR) 99.9% purity) carried by the $H_2$ flow were introduced into the quartz tube. After 10 min of reaction time, the furnace was cooled to room temperature, under $H_2$ flow, with various cooling rates from 40 ºC/min to 200 ºC/min. Experiment was also repeated for various growth temperatures and reaction times. The prepared samples were removed from the quartz tube and used directly for various spectroscopic and microscopic measurements.

*Instrumentation*

Confocal Raman measurements were done with a WiTec GmbH, Alpha-SNOM CRM 200 instrument having a 532 nm Nd:YAG laser as the excitation source. The excitation laser was focused using a 100X objective, and the signal was collected in a backscattering geometry and sent to the spectrometer through a multimode fibre. The effective scan range of the spectrometer was 0-9000cm$^{-1}$ (which amounts to a wavelength maximum of 1020.70 nm for 532 nm excitation), with the detection efficiency falling above 750 nm. A super-notch filter placed in the path of the signal effectively cuts off the excitation radiation. The signals were then dispersed using a grating of 150 grooves/mm and the dispersed light intensity was measured by a Peltier-cooled charge coupled device (CCD). Raman imaging was done using the same grating, with an integration time of 100 ms. Single spot spectra were also acquired with larger integration times. For improved resolution and to ascertain the peak positions, 1800 grooves/mm grating was used while acquiring single-spot spectra. The intensities of the desired portion of the spectra, collected over all of the pixels, were compared by Scan CTRL Spectroscopy Plus Version 1.32 software, to construct colour-coded images. Also, the image corresponding to various features of graphene, namely D, G and 2D, were filtered from the image using WiTec Project 3.2.

X-ray photoelectron spectroscopy (XPS) measurements were performed using an Omicron Nanotechnology ESCA Probe system with monochromatic Al Kα X-rays (energy of 1486.6 eV). X-ray power applied was 300 W. The survey spectra were collected at constant analyser energy (CAE) of 50 eV and the detailed regions were collected at a CAE of 20 eV and were averaged 15 times. Measurements were done at a base pressure of $10^{-9}$ mbar. The parent SS304 substrates alone were sputtered with $Ar^+$ ions (argon ions) at a base pressure of $10^{-6}$ mbar, to remove the oxide layers and other impurities from the surface of the sample prior to the XPS measurement.

Scanning electron microscopic (SEM) images and energy dispersive analysis of X-ray (EDAX) studies were done using a FEI QUANTA-200 ESEM with an EDAX system.

**Results and Discussion**

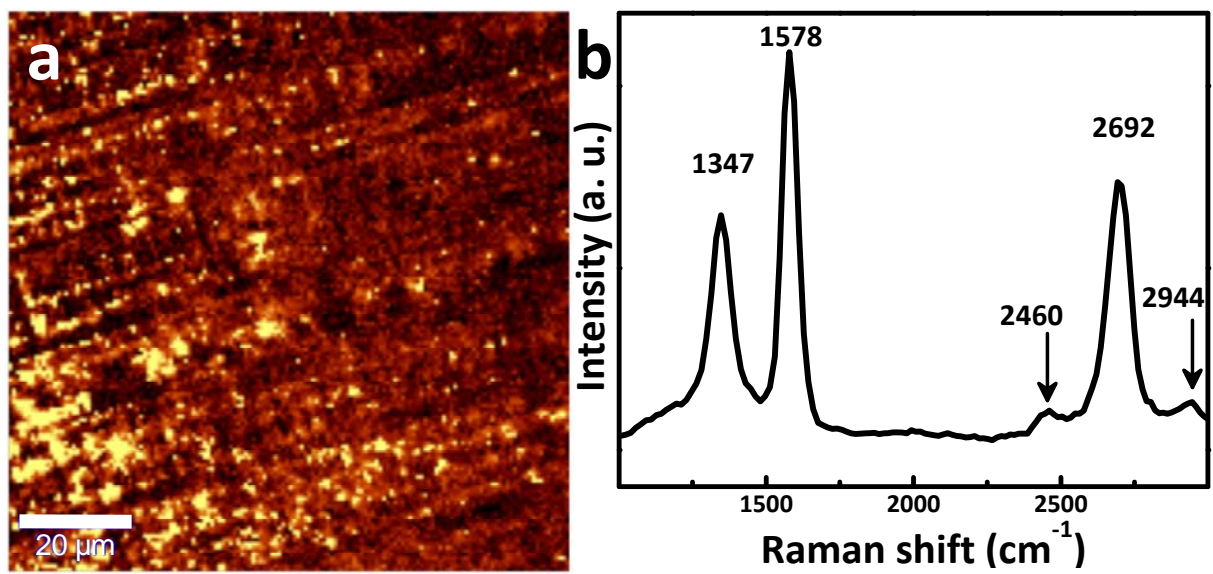

Figure 2. (a) Raman images corresponding to the graphene features (filtered for 1300-2750 $cm^{-1}$ range) from an area of 100 μm × 100 μm for a growth temperature, reaction time and cooling rate of 850 ºC, 10 min and 100 ºC/min, and (b) the corresponding average Raman spectrum from the 40,000 spectra, each of them corresponds to a pixel of the image.

Raman spectroscopy is established as the most accurate and easy tool to characterize graphene [2, 10-12], the number of layers and the presence of defects (quality) especially when the number of layers is less than 5 [17]. Figure 2 (a) is the Raman image from an area of 100 μm × 100 μm from the graphene grown sample for a growth temperature, reaction time and cooling rate of 850ºC, 10 min and 100ºC/min, respectively. Figure 2(b), an average Raman spectrum of the prepared sample shows the three most intense features at ~1347, ~1578 and ~2692 $cm^{-1}$. These are identified as the Raman fingerprints of graphene, namely of D, G and 2D (or known historically as G') peaks. The G peak is from the first-order Raman scattering process and is attributed to the doubly degenerate inplane longitudinal optic (iLO) and in-plane transverse optic (iTO) phonon modes ($E_{2g}$ symmetry) of $sp^2$ hybrid carbons at the Γ-point, whereas the D and 2D bands result from a second-order Raman processes. The D peak originates form a second order process involving one iTO phonon and a defect at the K-point and the 2D band is an overtone of the D peak involving two iTO phonons [18]. In addition, the D band is silent for infinite layers but becomes Raman active for a few layers with substantial number of defects [18]. Thus, the presence of D peak (~1347 cm$^{-1}$) suggests that the formed graphene is disordered and has inherent defects on it. These defects may include vacancies and strained hexagonal/non-hexagonal (pentagon or heptagon) distortions that lead to the non-uniformity, corrugation and twisting of the layers as shown in electron microscopic images (Supplementary information, figure S (1-2)). The low intensity peak at ~2460 cm$^{-1}$, known as G$^*$, is due to the intervalley double resonant Raman process similar to that of 2D band, but involving one LO and one iTO phonons [18]. Another weak feature at ~2944 cm$^{-1}$ is attributed to the combination of the D and G peaks [17]. Also, from figure 2 it is clear that the coverage of multi layer graphene on the SS substrate is pretty large. Our piezo-driven scan stage limits the confocal Raman imaging to a maximum area of 100 μm × 100 μm, though the graphene features were present throughout the substrate.

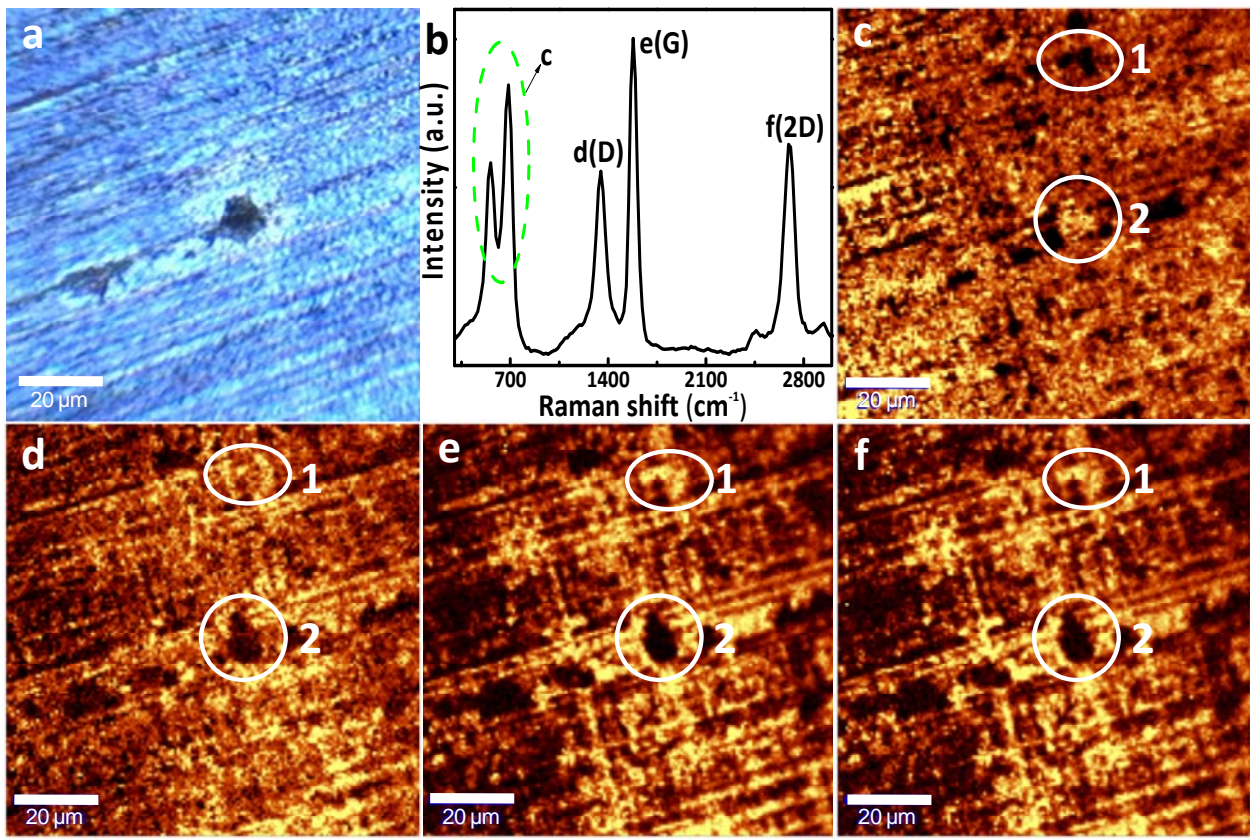

Figure 3. (a) Optical image; (b) single spot Raman spectrum; and Raman images filtered from the regions (c) 540-695 cm$^{-1}$, (d) D band (1300-1400 cm$^{-1}$), (e) G band (1550-1650 cm$^{-1}$) and (f) 2D band (2650-2730 cm$^{-1}$) for a 100 μm × 100 μm area.

Optical image and Raman maps corresponding to various regions of the Raman spectrum are shown in figure 3. Closer examination of the spectrum reveals additional features not attributed to graphene, at ~559 and ~686 cm$^{-1}$, marked collectively as in figure 3(b). The feature at ~559 cm$^{-1}$ is reported for $Cr_2O_3$, formed by the high temperature treatment of SS304, while the feature at ~ 686cm$^{-1}$ is reported for $MnCr_2O_4$ spinel [19-21]. Raman images filtered from spectral regions marked c, d (D band), e (G band) and f (2D band) on the Raman spectrum in figure 3(b) are shown in figure 3(c), 3(d), 3(e), and figure 3(f), respectively. It can be seen that the intensity of spot marked 1 in figure 3(c) is low (dark) and that of spot 2 is high, while the intensity of 1 is high and that of 2 is low in figures 3(d)-(f). Hence, it can be said that the peaks in regions c of figure 3(b) are mutually exclusive to those of other three images, namely d, e, and f, as established beyond doubt from the Raman maps. This suggests that whenever the features at regions c are prominent, graphene features (at regions d, e and f) are quenched and vice versa. Thus, depending on the compositional variation of SS304, we obtain Raman features of either metal oxides (such as $Cr_2O_3$, $MnCr_2O_4$) or graphene predominantly.

In figure 4 (a) Raman image for a 50 μm × 50 μm area is shown, on which different regions have been labeled 1, 2, 3, 4 and 5. Spectra (with 150 groves/mm grating) from these labeled regions of figure 4(a), along with that of graphite, were gathered in figure 4(b). The number of graphene layers is estimated from the intensities, shapes and positions of the G peak and 2D band. The Raman spectrum from the region 5 corresponds to the $Cr_2O_3$ and $MnCr_2O_4$ spinel. For further illustration, high resolution Raman spectrum from the regions 1, 2, 3, 4 and graphite with the 1800 groves/mm grating is presented in figure 5. The $I_G/I_{2D}$ ratio of the Raman spectrum from regions 1, 2, 3, 4 and graphite are ~0.73, ~0.91, ~0.95, ~1.21 and ~2.36 respectively, while corresponding position of the 2D band are ~2686 (full width at half maximum (FWHM) of ~32), ~2689 (~38), ~2693 (~42), ~2703 (~64) and ~2710 (~72) cm$^{-1}$. The Raman spectrum from region 1 shows the typical features of single-layer graphene such as $I_G/I_{2D}$ ratio ~0.7 and a symmetric 2D band centered around 2686 cm$^{-1}$, as shown in figure 5(b), with a full width at half maximum (FWHM) of ~32 cm$^{-1}$. This very low FWHM for the 2D band is comparable to the lowest reported [17]. Similarly, regions 2, 3 and 4 correspond to bi-, tri- and multi-layers of graphene. This is evident from the broadening of FWHM from ~32 to ~64 cm$^{-1}$, accompanied by the red shift in the position of the 2D band from ~2686 for single-layer to ~2703 cm$^{-1}$ for multi-layer graphene, as shown in figure 5(b). As the number of layers increases, the iTO phonon mode is getting branched resulting in the broadening of the 2D band. For more than 5 layers, the Raman spectrum of graphene becomes hardly distinguishable from that of graphite [18]. The positions of the D and G bands remain the same for different number of layers as evident from figure 5(a). The ratio of the intensities of the G band to the 2D band is more than the expected ratio of 0.25 for graphene transferred onto SiO$_2$ substrate. Such enhancement in the ratio suggests that the grown graphene is doped with electrons from the substrate [22]. This is evident from the high intensity of D band and the position of the 2D band, which for an undoped graphene is ~2680 cm$^{-1}$.

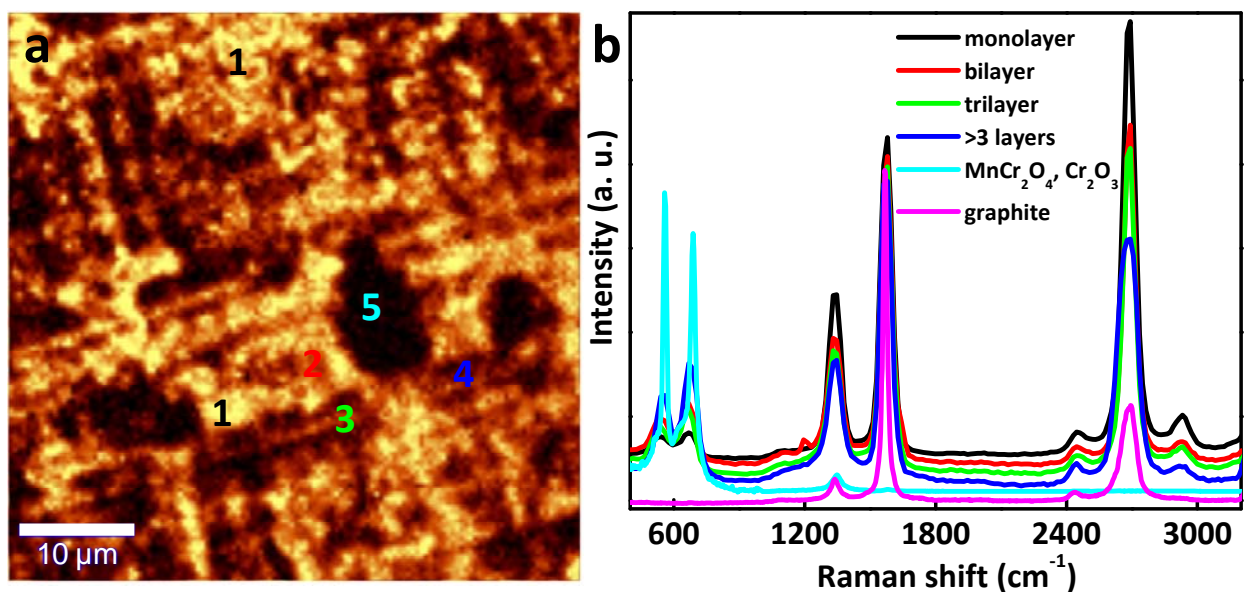

Figure 4(a). Raman image (filtered for 2650-2730 cm$^{-1}$) showing areas of (1) single-, (2) bi-, (3) tri- and (4) multi-layers of graphene and the region marked 5 contains $Cr_2O_3$ and $MnCr_2O_4$ spinel; (b) the corresponding spectra with 150 groves/mm grating, after normalizing with respect to G peak, for a 50 μm × 50 μm area.

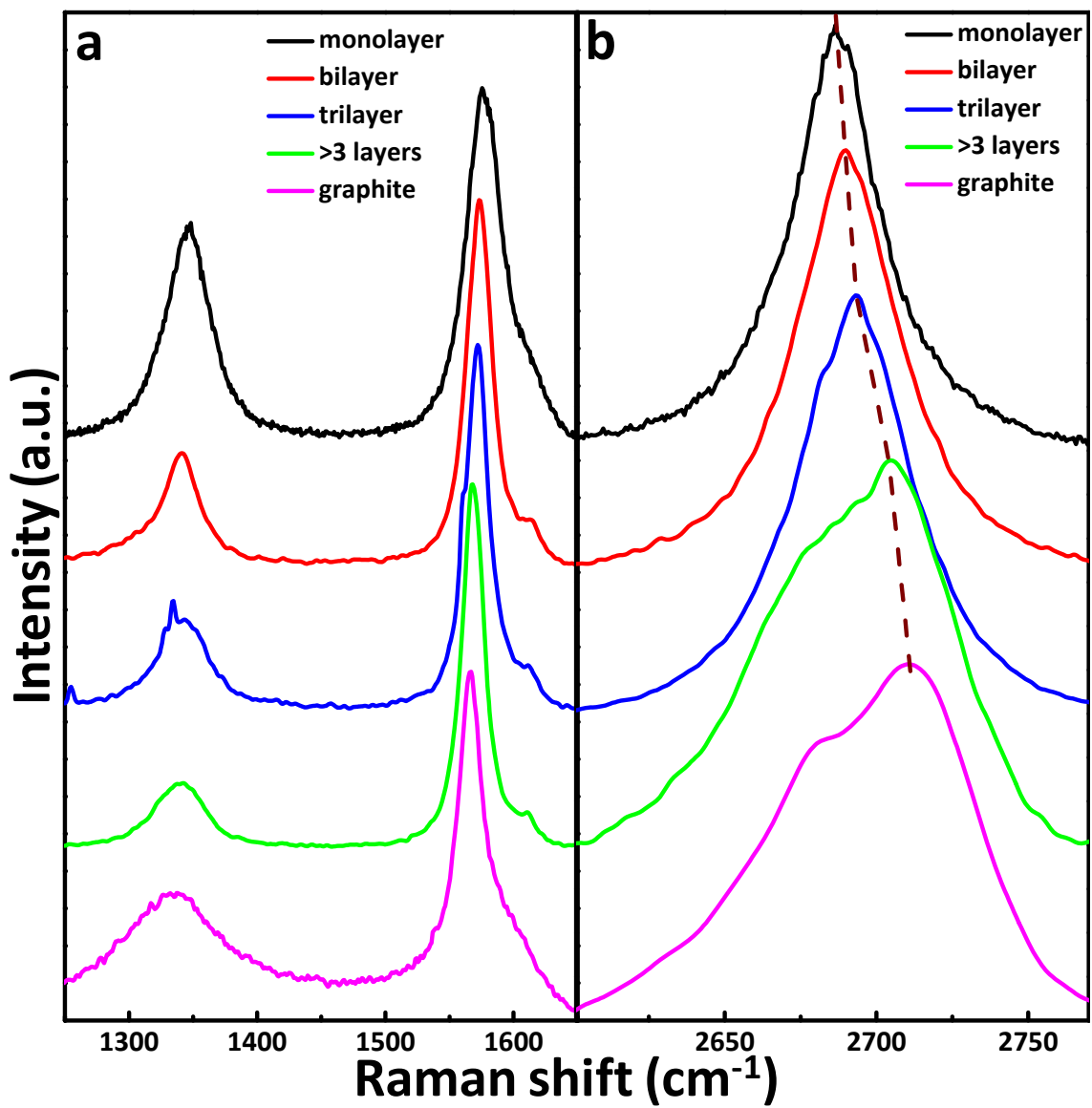

Figure 5. Evolution of the Raman spectra with number of layers, with 1800 groves/mm grating; (a) the D band is positioned at 1347 cm$^{-1}$ and the G band is positioned at 1578 cm$^{-1}$, without any considerable change with increase in the number of layers and (b) the 2D band whose position undergoes a blue shift from 2686 cm$^{-1}$ for graphene to

2703 cm$^{-1}$ for multi-layer graphene along with an increase in FWHM from 32 to 64 cm$^{-1}$. The features corresponding to graphite is also included for comparison.

The graphene growth is found to be highly sensitive to various parameters, the most important being the growth temperature and the cooling rate. To explore the effect of both on the graphene growth, we have repeated the experiment at various growth temperatures and cooling rates (Supplementary information, figures S3 & S4). The optimum growth temperature to obtain maximum graphene coverage was found to be 850 ºC. At this temperature, the cooling rate was varied from ~40–200 ºC/min. At very high cooling rates ($\geq$ 140 $^0$C/minute) no signatures of graphene was observed. At low cooling rates, Raman analysis of the samples showed features similar to that of graphite. Our analysis suggests that sedimentation of carbon to form graphene was optimum at a cooling rate of ~100 ºC/min. We have also conducted experiments with the SS304 itself, without any carbon source, as the carbon in SS304 itself can act as the source of carbon. However this yielded graphene with a very low coverage (Supplementary information, figure S (5)). The CVD method is economical and we suggest that the total time taken for the experiment can be reduced if the time required for flushing is reduced by employing a vacuum pump.

To understand the origin of the observed Raman features (mainly peaks at 559 & 686 cm$^{-1}$) and chemical nature of the surface, XPS and SEM-EDAX investigations were carried out. Figure 6 shows the XPS spectra of graphene grown SS304 (GS) and the parent SS304 (PS) substrates. In the survey spectra, figure 6(a), of both samples show characteristic elements such as C, Cr, Fe, Mn, Ni and Si along with O. C 1s intensity of GS was higher than that of PS, same is the case with Cr 2p and Mn 2p. However, the intensity of Fe 2p and Ni 2p were found to be very weak for GS; while they are prominent for PS. The GS surface showed characteristic features of oxides of Cr and the other constituent metals. For a detailed analysis, similar regions in the resolved and normalized spectra of both the PS and GS substrates were compared. Presence of elements in various chemical forms has been summarized in Table 1 and briefly discussed below.

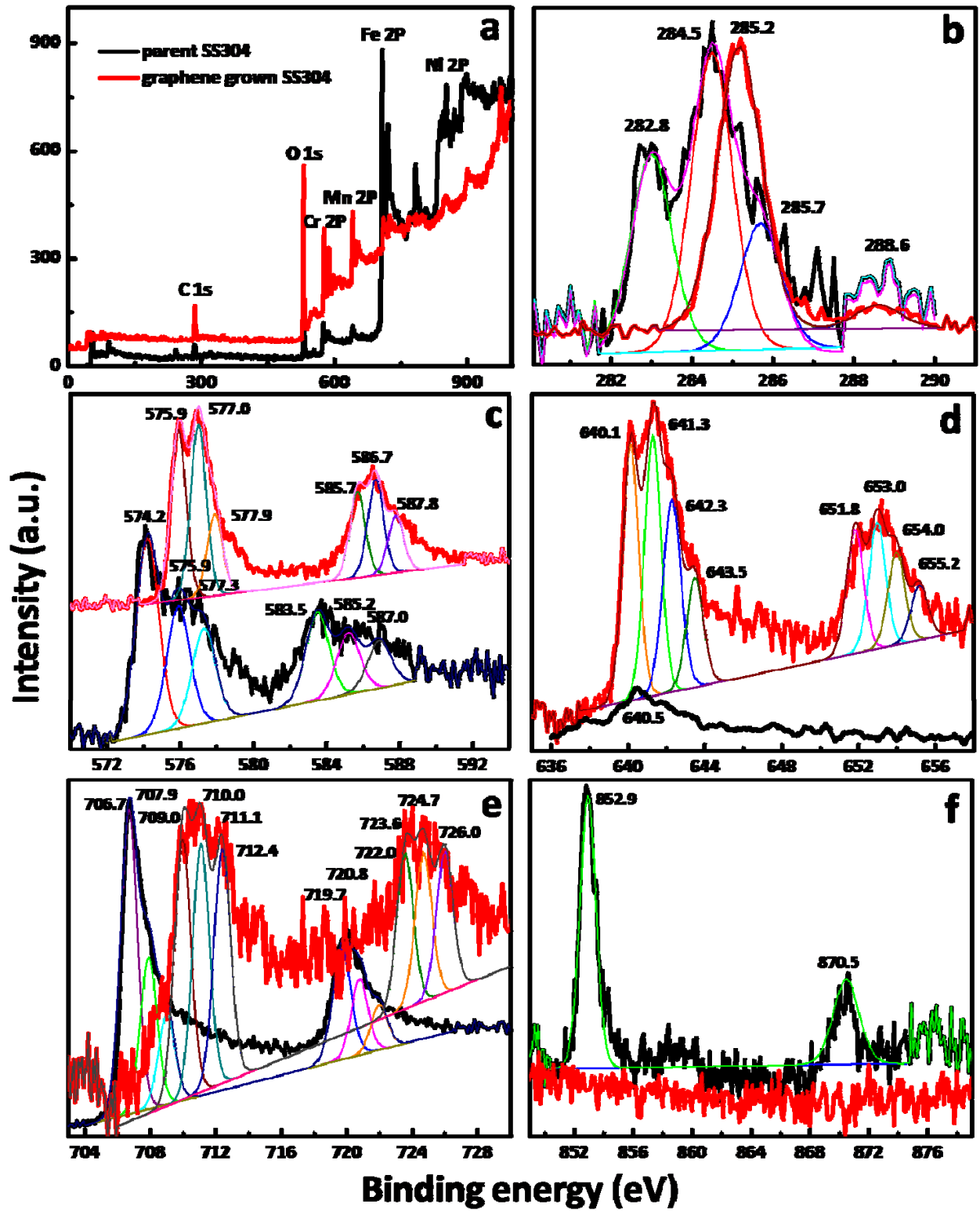

Figure 6. (a) The X-ray photoelectron survey spectra and the regions corresponding to (b) C 1s; (c) Cr 2p; (d) Mn 2p; (e) Fe 2p; and (f) Ni 2p for graphene grown and parent SS304 substrates, measured with monochromatic Al Kα X-rays. The various components were fitted. Spectrum for GS was translated vertically upwards in 6 (c) for better clarity.

The deconvoluted C 1s spectrum of PS, figure 6(b), shows three distinct Gaussian features indicating the presence of transition metal carbide(s), adsorbed carbon and organic impurities on the surface [23]; while the C 1s spectrum of GS shows two features with the prominent feature corresponds to covalent $sp^2$ hybridized carbon [22, 23], confirming the presence of graphene, i.e., carbon in zero-valent state, where as the other component shows the presence of carboxyl functional groups on the formed graphene. The Cr 2p spectrum of GS shows three features, indicating the presence of chromium carbide, $MnCr_2O_4$ spinel and $Cr_2O_3$ [22]. These are the two species ($MnCr_2O_4$ spinel and $Cr_2O_3$) which give rise to the distinct Raman features at 559 & 686 cm$^{-1}$ respectively.

Table 1. Various chemical species present on the surfaces of PS and GS

| Element | Prominent feature | Sample type | Deconvoluted XPS components (eV) | Possible chemical species |
|---|---|---|---|---|
| C | C 1s | PS | 282.8, 284.5, 285.7, 288.6 | Metal carbide, adsorbed carbon, organic impurities & carboxyl group |
| | | GS | 285.2, 288.6 | $sp^2$ hybrid carbon & carboxyl |
| Cr | Cr 2p$_{3/2}$ | PS | 574.2, 575.9, 577.3 | Metallic chromium, chromium carbide & CrO |
| | | GS | 575.9, 577, 577.9 | Chromium carbide, $MnCr_2O_4$ spinel & $Cr_2O_3$ |
| Mn | Mn 2p$_{3/2}$ | PS | 640.5 | MnO |
| | | GS | 640.1, 641.3, 642.3, 643.5 | $Mn_3C$, $Mn_2O_3$, $MnO_2$ & $MnCr_2O_4$ |
| Fe | Fe 2p$_{3/2}$ | PS | 706.7, 707.9, 709 | Metallic Fe, $Fe_3C$ & FeO |
| | | GS | 710, 711.1, 712.4 | FeO, $Fe_3O_4$ & $Fe_2O_3$ |
| Ni | Ni 2p$_{3/2}$ | PS | 852.9 | Metallic Ni |
| | | GS | --- | |

Mn 2p spectrum, figure 6(d), of PS shows the presence of MnO, while the same for GS shows a 2p$_{3/2}$ feature with three components indicating the presence of $Mn_3C$, $Mn_2O_3$ and $MnO_2$, in accordance with the $MnCr_2O_4$ phase [22]. The assignment of the surface species correspond to the most probable cases considering the available data. The data suggest that during the process of heating, Mn is migrating to the SS surface in the form of $MnCr_2O_4$ and other oxides Mn and hence increasing the Mn concentration at the surface [25]. This is also evident from Mn $2p_{3/2}$ intensity of GS which appears to be negligible in case of PS. Fe $2P_{3/2}$ spectra, figure 6(e), of both the PS and GS substrates show a three component feature similar to those of the other transition metals. Spectral features suggest that GS is composed of oxides of iron: $Fe_3O_4$, $Fe_2O_3$ and FeO. FeO present on the PS acts as the source of oxygen for the oxidation of dissolved carbon to form CO, which then disproportionate to graphite and $CO_2$ upon direct heating of SS. In the case of alcohol as an additional carbon source, CO required for the disproportionation can also be due to the pyrolysis of ethanol. However, the presence of Cr or Mn species at the GS surface favors the formation of $MnCr_2O_4$ phase or their respective metal carbides by reacting with the reduced carbon, thereby preventing the formation of graphene. Therefore, cooling rate is crucial to have an optimum growth of single- or few-layered graphene. This is evident from the mutual exclusive nature of peaks as established in figure 3. The Ni 2p region of the survey spectrum, figure 6(a), of both PS and GS suggest the presence of metallic Ni in both [23], except that the surface concentration of Ni is very low to be detected by monochromated XPS, as evident from figure 6(f).

Thus it is evident that the surface of GS contains several oxides of transition metals; oxides of Cr & Mn in particular, namely $Cr_2O_3$ and $MnCr_2O_4$ spinel. These are in good agreement with the Raman data and are responsible for the two Raman peaks labeled as 'c' in figure 3(b). These species hinder graphene growth, as evident from the mutual exclusive nature in Raman images (figure 3(c)-(f)). The Raman and the XPS analyses along with the vibrational characterization of untreated commercial PS and GS, conclusively establish that presence of excess Cr and Mn retard the growth of graphene, while FeO helps in forming the reduced $sp^2$ hybridized carbon and hence graphene. The role of a particular growth temperature and cooling rate on the formation of graphene can be justified by considering several competing events which are optimized at the given set of parameters.

**Conclusions**

The present study has shown that it is possible to grow highly organized graphene samples by a simple thermal CVD method, using a laboratory assembled set up. Our method of synthesis using hot wall split tube furnace, relatively cheap substrate (SS304) and carbon source (ethanol) proves to be economical for the synthesis of graphene. The presence of single-, bi- and tri-layers of graphene over a large area is established by Raman spectroscopy and microscopy. Raman spectroscopic data in combination with XPS analysis suggest that some oxide species of Mn & Cr along with a spinel structured compound $MnCr_2O_4$ play a key role in the non homogeneity of the formed graphene. The detailed knowledge of the chemical species present on the surface can be useful in preparing graphene based catalytic surfaces on SS304. Present studies may be extended to create an iron based catalyst system for single-layer graphene growth.

## Acknowledgments

We thank the Department of Science and Technology, Government of India for constantly supporting our research program on nanomaterials. RJ thanks UGC for a Junior Research Fellowship. AA expresses his gratitude to Tirupatireddy K Keshav, Founder Chairman of IUWS for introducing him to the field of Nanoscience & technology.

## Supplementary information

Raman spectra of graphene grown under different growth conditions along with the SEM and EDAX analysis of the graphene.

# Supplementary information

# Single- and few-layer graphene growth on stainless steel substrates by direct thermal chemical vapor deposition

Robin John,[a§] A Ashokreddy[b§] C Vijayan and T Pradeep[b*]

[a] Department of Physics, Indian Institute of Technology Madras, Chennai – 600 036, India.

[b] DST Unit of Nanoscience (DST UNS), Department of Chemistry, Indian Institute of Technology Madras, Chennai – 600 036, India.

[§] These authors contributed equally to this work.

[*] Corresponding author. Fax: + 91-44 2257-0545. E-mail address: pradeep@iitm.ac.in (T. Pradeep)

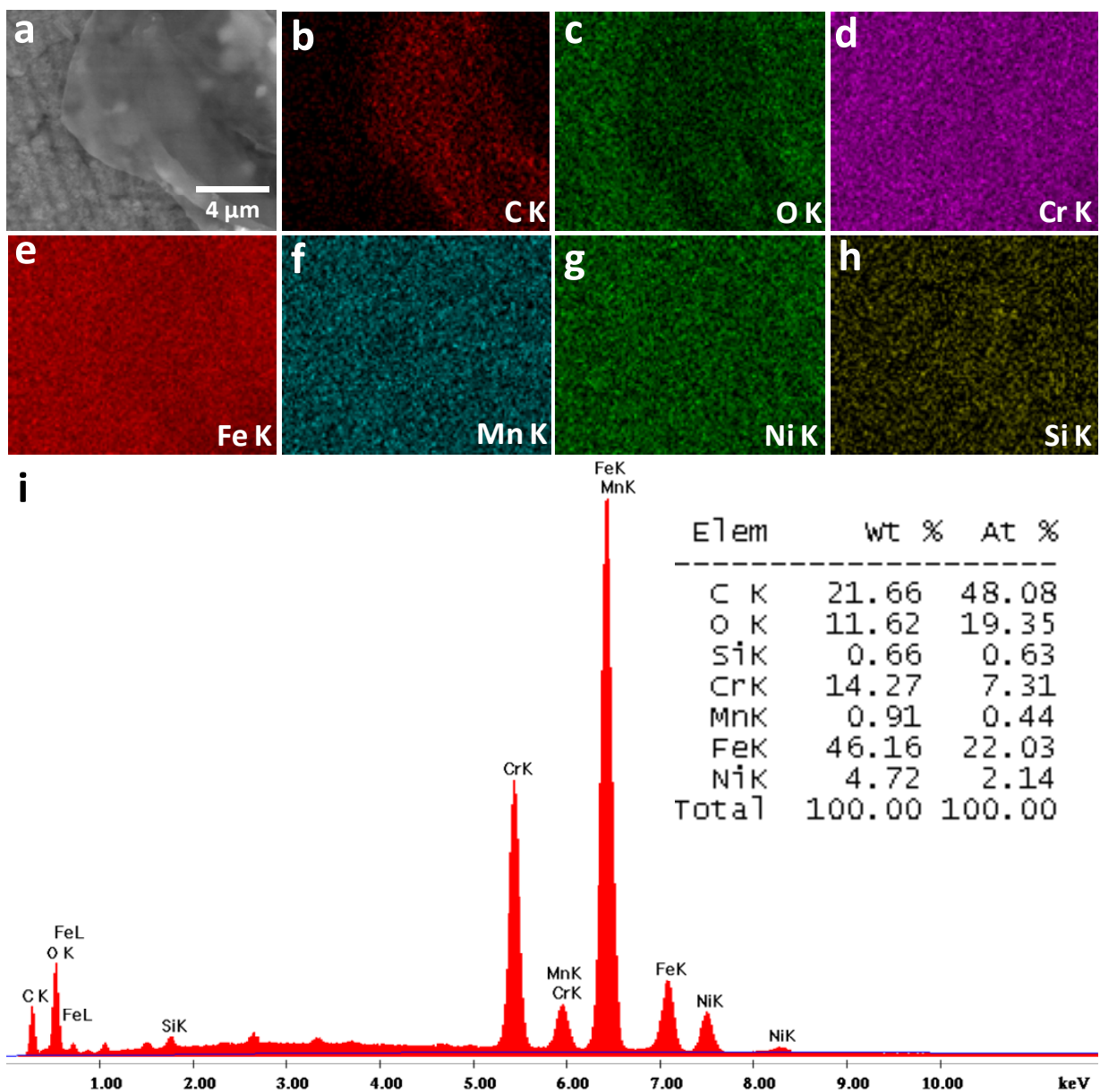

Figure S1. SEM and EDAX analysis of grown graphene at a growth temperature and cooling rate of 850ºC and 10 min respectively, with ethanol as carbon source. (a) SEM image; EDAX images collected using (b) C K, (c) O K, (d) Cr K, (e) Fe K, (f) Mn K, (g) Ni K, (h) Si K; and (i) the corresponding EDAX spectrum showing elemental composition.

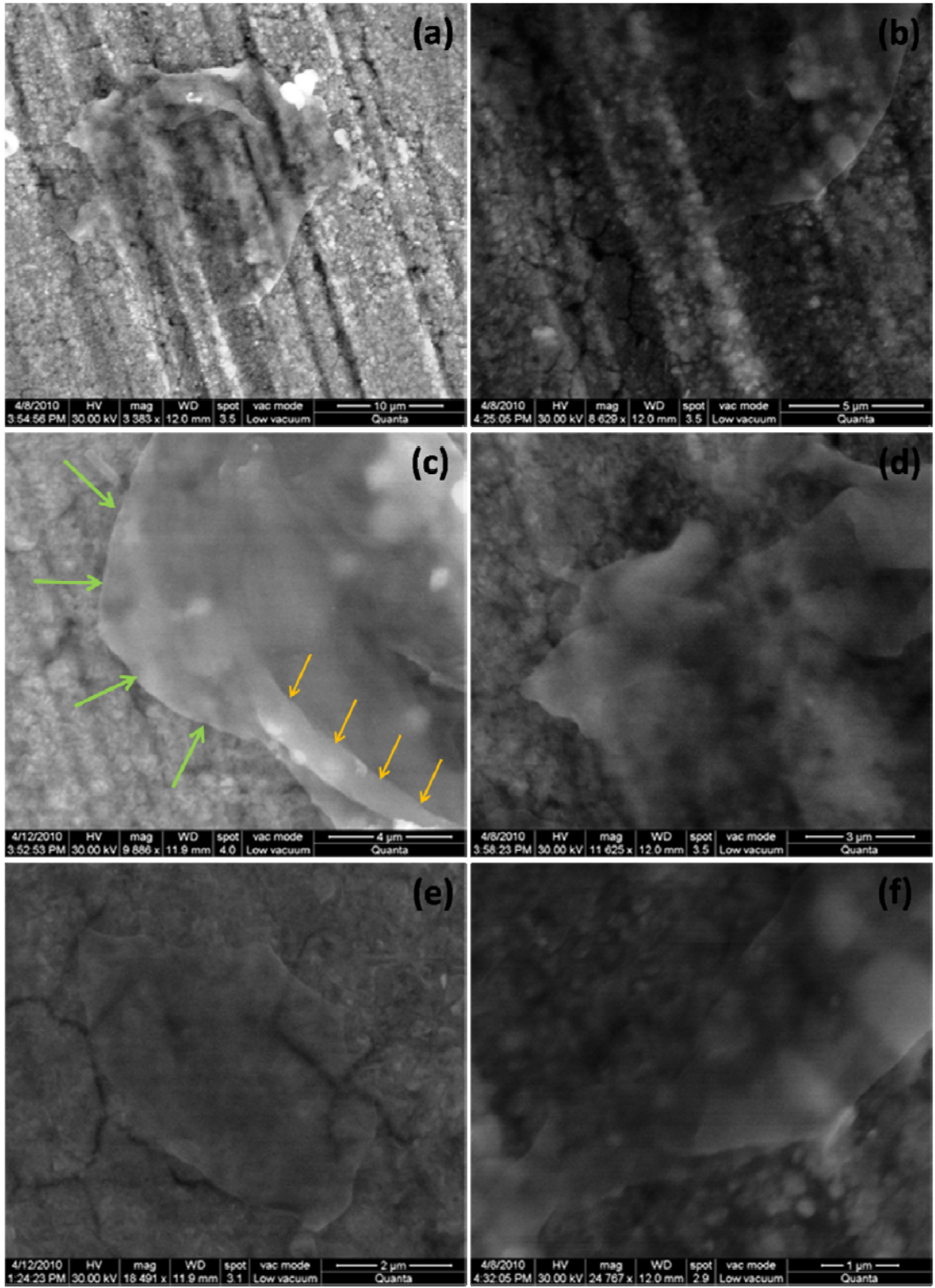

Figure S2(a-f). SEM images of multi-layer graphene under different magnifications; (c) Multi-layer features are evident from the images with the edges marked by green arrows and corrugations on the layers are marked by orange colored arrows.

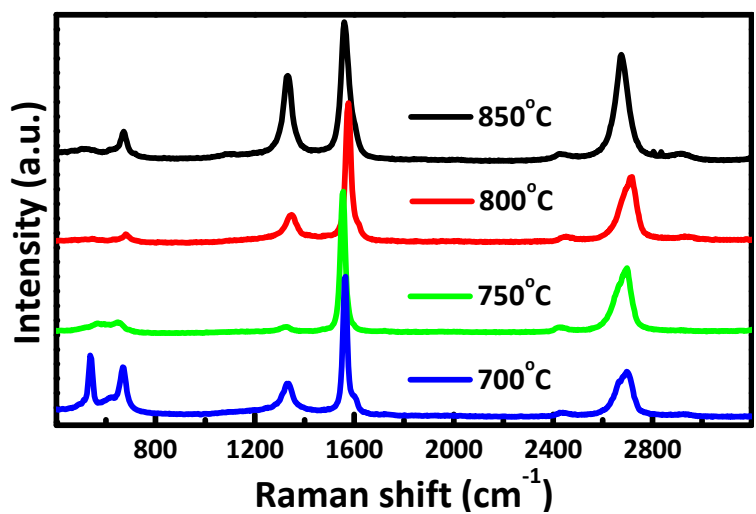

Figure S3. Raman spectra obtained at various temperatures with ethanol as the carbon source. Features indicate multi-layers of graphene.

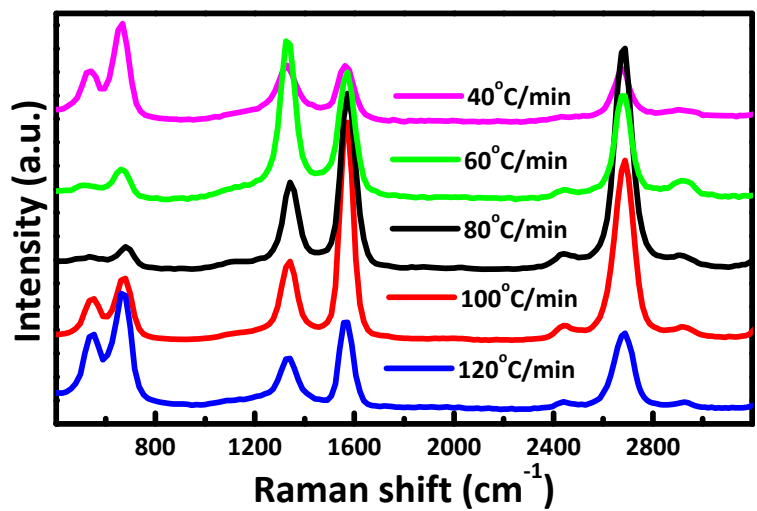

Figure S4. Raman spectra obtained for various cooling rates with ethanol as the carbon source and a reaction time of 10 minutes, at 850ºC.

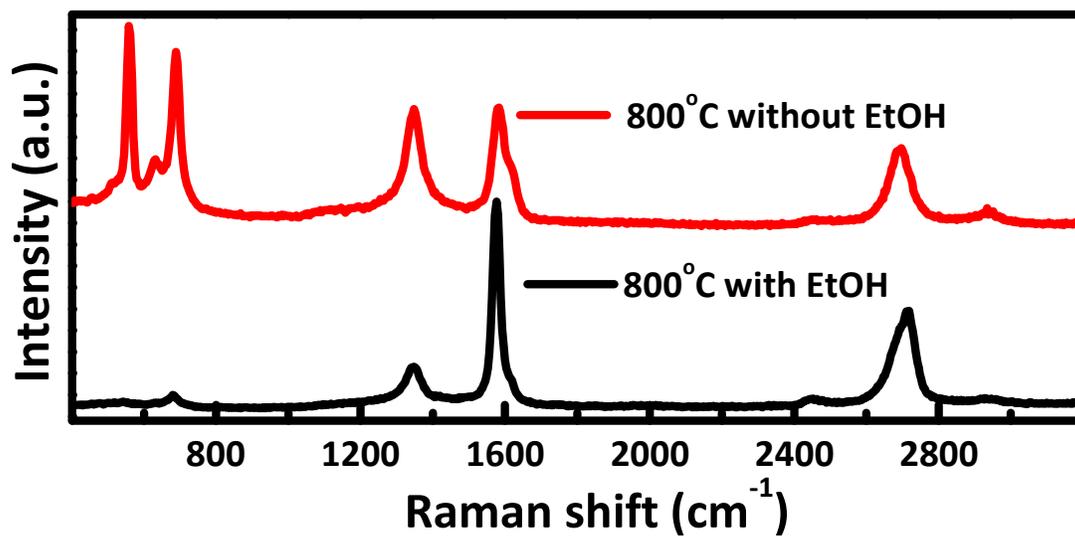

Figure S5. Raman spectra obtained with and without ethanol as the carbon source, at 850ºC at a cooling rate of 100 $^0$C/minute.